\documentstyle[graphicx,multicol,prl,aps]{revtex}

\begin{document}
\pagestyle{empty}
\begin{multicols}{2}
\narrowtext
\parskip=0cm
\noindent
{\bf Comment on ``Triviality of the Ground State Structure in Ising
Spin Glasses''}

In a recent, very interesting paper, Palassini and Young \cite{PALYOU}
have shown that it is possible to get useful information about the
nature of the low $T$ phase of $3D$ Ising spin glasses by studying the
behavior of the ground state (GS) after changing the boundary
conditions (BC) of an $L^{3}$ lattice system from periodic ($P$) to
anti-periodic ($AP$).  They analyze GS obtained with the same
realization of the quenched disorder and different BC.

Let $P(M,L)$ be the probability that the spins in an $M^{3}$ cube
remain in the same configuration (apart from a full reversal, due to
the global $Z_2$ symmetry at zero magnetic field) when we change BC
from $P$ to $AP$ \cite{WINDOW}.  The behavior of $P(M,L)$ when $L$
goes to infinity is very important from the theoretical point of view.
In the droplet model (DM) \cite{DROPLE} $P(M,L)\propto L^{-\lambda}$,
where $\lambda \equiv D-D_{s}$, $D$ is the space dimension and $D_{s}$
is the fractal dimension of the interface, while in the usual form of
the Replica Symmetry Breaking (RSB) approach \cite{RSB} $P(M,L) \to
A(M)$, where $A(M)$ is a non zero function (i.e. the interface is
space filling).  In \cite{PALYOU} it is shown that the data for
$P(2,L)$ can be well fitted by a power law with a non-zero $\lambda$
(as suggested by the DM), although they can also be fitted as
$a+bL^{-1}+cL^{-2}$, with a non zero $a$.  In this comment we point
out that one can better discriminate among the DM and the RSB approach
if one extends their analysis analyzing the value of additional
quantities. At this end we have computed the GS in systems with side
up to $L=12$ (with Gaussian disorder) and we have compared the GS
obtained with $P$ and $AP$ BC.

If in the large volume limit the interface is a {\em homogeneous}
fractal that can be characterized by a single fractal dimension
(i.e. if it has not a multi-fractal behavior), and if the relation
$\lambda=D-D_{s}$ is correct, the probability that the interface does
not intersect a region $\cal R$, whose size is proportional to the system
size, goes to a limit which is a non-trivial function of the shape of
the region $\cal R$.  If the interface is space filling, such a probability
always goes to zero.  This argument implies that under the previous
assumptions in the DM (for large $L$) $P(M,L)\propto g(ML^{-1})$.

\begin{figure}
\centering
\includegraphics[width=0.27\textwidth,angle=270]{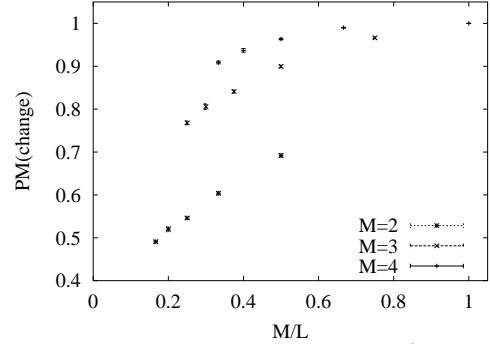}
\caption[a]{$P(M,L)$ versus $ML^{-1}$.  
\protect\label{fig1} }
\end{figure}

\begin{figure}
\centering 
\includegraphics[width=0.27\textwidth,angle=270]{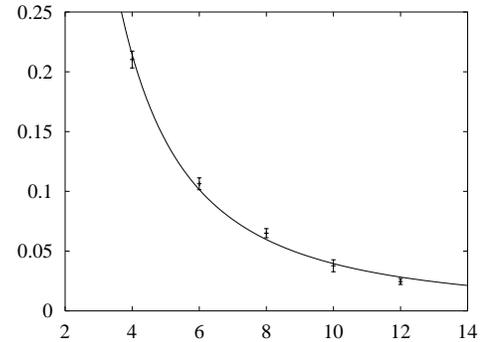}
\caption[a]{$P_L$ as defined in the text versus $L$. 
\protect\label{fig2} }
\end{figure}

We plot in figure (\ref{fig1}) our results for boxes of size $M$ $=$
$2$, $3$, $4$ versus $ML^{-1}$.  The data are very far from collapsing
on a single universal curve (they are consistent with a smooth
behavior in $L^{-1}$, and are well fitted by a second order polynomial
in $L^{-1}$).  Stronger hints are obtained if we consider the
probability $P_{L}$ of finding that a full $y-z$ plane of $L^{2}$
spins does not hit the interface when we go from $P$ to $AP$ in the
$x$ direction.  This corresponds in the previous argument to consider
a region $\cal R$ of size $L \times L \times 1$. In figure
(\ref{fig2}) we plot $P_L$ versus $L$.  $P_L$ can be roughly fitted as
$L^{-\gamma}$, with a relative large value of $\gamma$ (i.e $\gamma
\approx 1.5-2.0$).  In other words the probability that the interface
hits $\cal R$ goes to one (or to a value very close to one) when the
volume goes to infinity.

We have shown that extending the innovative analysis of Palassini and
Young to a larger set of observables, one finds serious problems with
the usual DM interpretation: the most natural scenario is based on the
fact that the interface is space filling as predicted by the RSB
approach. Other possibilities like the presence of very strong
corrections to the scaling, or that the relation $\lambda=D-D_{s}$ is
not valid and/or the interface is multi-fractal, are less plausible.
We are grateful to M. Palassini and P. Young for pointing out an error
in the interpretation of the data in a first version of this comment
and for a very useful correspondence.

\noindent 
E. Marinari and G. Parisi,
{\small   Universit\`a di Roma {\em La Sapienza}}

\end{multicols}
\end{document}